%% file: root.tex
\documentclass{ifacconf}

\usepackage{graphicx}      
\usepackage{natbib}        
\usepackage{caption}
\usepackage{subcaption}
\usepackage{enumitem}
\usepackage{balance}
\usepackage{amsfonts}
\usepackage{amssymb}
\usepackage{amsmath}
\usepackage{float}
\usepackage{xurl}
\usepackage{booktabs}

\begin{document}
\begin{frontmatter}

\title{Passive Fault-Tolerant Antenna Stabilisation for Ship-to-Shore Communication\thanksref{footnoteinfo}}

\thanks[footnoteinfo]{The present work is supported by the fund File No. 8090-00063B and is part of SLGREEN, carried out under the ShippingLab platform. ShippingLab is supported by Innovation Fund Denmark, the Danish Maritime Fund, DS Norden, and the Lauritzen Foundation. The authors much appreciate this support.}

\author[First]{Antoine Thibault Vi{\'e}}
\author[First]{Mikkel Reinholdt Hansen}
\author[First]{Roberto Galeazzi}
\author[First]{Dimitrios Papageorgiou}

\address[First]{Technical University of Denmark, Department of Electrical and
Photonics Engineering, Elektrovej 326, 2800 Kgs Lyngby, Denmark
(e-mails: \{anthv, roga, dimpa\}@dtu.dk, mikkel-r-hansen@live.dk).}

\begin{abstract}
Directional ship-to-shore radio links require continuous antenna alignment with the shore station, while vessel heave induces variations in antenna height above the sea surface, thereby altering the multipath interference governing the received power. In conventional stabilised gimbals, attitude compensation is provided, whereas fault tolerance is rarely considered. To address these limitations, a passive fault-tolerant control scheme is proposed, whereby the antenna is mounted on a ship-mounted six-degree-of-freedom manipulator, enabling simultaneous regulation alignment and antenna height. Through feedback linearisation, the manipulator dynamics are decoupled into perturbed integrators, each controlled by a multi-barrier adaptive super-twisting sliding-mode controller. The nested barriers are associated with the performance regions of fault-tolerant control, such that inter-sample barrier crossings under discrete-time implementation lead to graceful performance degradation rather than loss of confinement. Under simulated conditions, superior pointing and heave accuracy is achieved compared with single-barrier designs, for which either precision or confinement is compromised.
\end{abstract}


\end{frontmatter}

\section{Introduction} \label{sec:intro}
    \input{intro.tex}

\section{Problem formulation} \label{sec:problemFormulation}

\input{problemFormulation.tex}

\section{Resilient ship-to-base communication} \label{sec:controlDesign}
    \input{controlDesign.tex}


\section{Simulations} \label{sec:rerults}
    \input{results.tex}

\section{Conclusions and future work}
    \input{conclusions.tex} \label{sec:conclusions}


\balance
\bibliography{resilientCommSafeProc2026}  


\end{document}

%% file: intro.tex




In maritime communication systems, accurate antenna alignment is essential for maintaining signal integrity. Rigorous pointing specifications have been documented, wherein root-mean-square (RMS) tracking errors are constrained to thresholds as minimal as $0.2^\circ$~\citep{wang_2021_attitude}. Achieving such precision in marine environments is challenging by continuous vessel dynamics. Furthermore, additional disturbances are introduced by wind loading and structural compliance. In industrial practice, antenna stabilisation is commonly achieved using mechanically stabilised platforms, such as motorised gimbals equipped with inertial sensors and feedback control. Passive mechanisms, including balanced gimbals and gyroscopic stabilisation, have also been employed to attenuate the effects of platform motion. Although these methodologies mitigate angular disturbances and preserve the requisite line-of-sight (LoS) alignment with a base station or satellite, the optimization of antenna pointing accuracy remains their primary functional objective.


However, maintaining angular alignment alone does not ensure optimal communication channel conditions. In particular, vessel heave changes the antenna height relative to the sea surface, thereby modifying the propagation geometry and the phase difference between the direct LoS path and sea-surface reflections \citep{wang_2018_wireless}. Depending on the operating conditions, these variations can induce destructive multipath interference and substantial fluctuations in received signal power~\citep{vie2025extended}. 
Consequently, a system that maintains accurate angular alignment may still experience severe channel degradation as the vessel moves vertically. These considerations motivate extending conventional antenna stabilisation beyond attitude compensation to account explicitly for vertical motion. By actively regulating antenna height, the stabilisation system can potentially mitigate heave-induced variations in maritime propagation conditions, complementing conventional pointing control.

Sliding mode control (SMC) is well known for its robustness to matched uncertainties and external disturbances~\citep{utkin_1977_variable}. Among higher-order methods, the Super-Twisting Sliding Mode Controller (STSMC) provides finite-time convergence with a continuous control output, reducing chattering in electromechanical applications~\citep{shtessel2010super, moreno_2012_lyapunov, moreno_2014_strict}. The inherent robustness of SMC methods against--and often their insensitivity to--bounded and Lipschitz perturbations, position them as suitable \emph{passive fault-tolerant control} solutions \citep{estrada2024passive} under the consideration that faults manifest as perturbations in the system dynamics. However, classical STSMC relies on fixed gains designed to dominate worst-case disturbances. Under highly variable maritime conditions, such bounds are difficult to estimate, leading to excessive control effort during calm periods or insufficient rejection of severe perturbations~\citep{eisenzopf2021adaptive, papageorgiou2022behaviour}.

Adaptive STSMC addresses this limitation through online gain adjustment~\citep{shtessel2010super, moreno_2015_adaptive, utkin_2013_adaptive}. Nevertheless, convergence is often guaranteed only to a neighbourhood of the sliding manifold, with its size and reaching time dependent on the unknown disturbance derivative bound~\citep{obeid_2018_barrier}. Moreover, excessive gains may be generated when disturbance bounds are unavailable, increasing control effort and degrading performance~\citep{gonzalez2024final}.

Barrier-function-based adaptive SMC has therefore been investigated to enforce prescribed performance without prior knowledge of disturbance bounds~\citep{obeid_2018_barrier}. Extensions to STSMC and higher-order sliding-mode controllers have demonstrated improved robustness~\citep{obeid_2020_barrier, laghrouche_2021_barrier}, while signed-power and bi-powered formulations have enabled less conservative gain tuning near performance boundaries~\citep{cruz_ancona_2021_barrier, gonzalez2024final, ovalle_2025_noise}. However, discrete-time implementations remain susceptible to inter-sample violations of barrier constraints. To address this limitation, a multi-layer STSMC framework was presented in~\citep{vie_MB}, which introduced $N$ performance layers using nested barriers, allowing performance degradation to be managed across admissible layers.

This work presents a multi-layer STSMC-based passive fault tolerant control solution for antenna alignment and stabilisation to facilitate resilient ship-to-base communication. The proposed approach is based on employing a 6 degrees of freedom (DOF) robotic manipulator that directs the attached antenna to a predefined reference pose. This ensures alignment of the antenna to the shore communication station.

The contributions of this work are:
\begin{itemize}
    \item Application of the multi-layer barrier-based adaptive SMC algorithm introduced in~\citep{vie_MB} to a multiple-inputs-multiple-outputs (MIMO) system by employing independent joint control. This is possible due to the property of the controller to handle residual cross-coupling terms as perturbations.
    \item Explicit mapping of the multiple layers of the proposed algorithm to admissible domains of operation in the joint space, i.e. nominal, degraded performance and unsafe operation.
    \item Implementation of the proposed control scheme in discrete time and demonstration of its fault-tolerant capabilities in simulation under faults and realistic uncertain perturbations, such as waves, wind and increased robot joint friction.
    \item Quantitative evaluation of the proposed approach based on antenna-pointing accuracy and heave stabilisation, as well as overall performance improvement.
\end{itemize}


The remainder of this paper is organised as follows: Section~\ref{sec:problemFormulation} formulates the problem. Section~\ref{sec:controlDesign} presents the barrier-based adaptive STSMC, control architecture, and underlying assumptions, followed by the derivation of the control laws. Simulation results are presented in Section~\ref{sec:rerults}, and conclusions and future research directions are provided in Section~\ref{sec:conclusions}.

%% file: problemformulation.tex

\subsection{System description}

The system under study is a 6-DOF serial manipulator mounted on a ship as illustrated in Figure~\ref{fig:problem_visualisation}. Its dynamics is described by
\begin{equation}
    D(q)\ddot{q} + C(q,\dot{q})\dot{q} + G(q) = \tau+\tau_d
    \label{eq:mani_dynamics}
\end{equation}
where \(q,\dot q,\ddot q\in\mathbb{R}^6\) denote the joint positions, velocities, and accelerations, respectively. \(D(q)\in\mathbb{R}^{6\times6}\) is the inertia matrix, \(C(q,\dot q)\in\mathbb{R}^{6\times6}\) represents the Coriolis and centrifugal effects, \(G(q)\in\mathbb{R}^6\) denotes the gravitational torque vector, \(\tau\in\mathbb{R}^6\) is the control torque vector and \(\tau_d\in\mathbb{R}^6\) represents the lumped environmental disturbance, defined as
\begin{equation}
    \tau_d \triangleq \tau_{\mathrm{wave}} + \tau_{\mathrm{wind}} + \tau_{\mathrm{fric}}
    \label{eq:disturbance_decomposition}
\end{equation}
where the individual terms represent the effects of vessel-induced wave motion, wind loading, and friction, respectively.

\subsection{Problem formulation}
\begin{figure}[H]
  \centering
  \includegraphics[width=\linewidth]{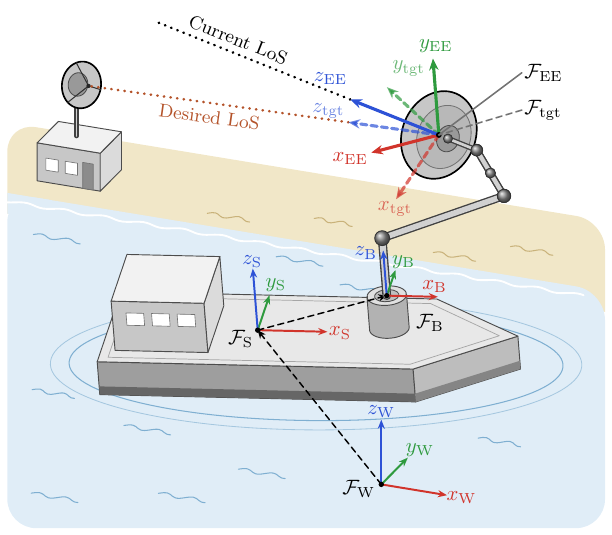}
  \caption{Visualisation of the antenna stabilisation problem.}
  \label{fig:problem_visualisation}
\end{figure}

To maintain resilient communication, the end-effector frame $\mathcal{F}_\mathrm{EE}$ must track a desired target frame $\mathcal{F}_\mathrm{tgt}$, as illustrated in Figure~\ref{fig:problem_visualisation}.

The target pose is provided by an upstream communication link optimisation scheme and expressed in the world frame $\mathcal{F}_\mathrm{W}$. The manipulator base frame $\mathcal{F}_\mathrm{B}$ is rigidly attached to the vessel frame $\mathcal{F}_\mathrm{S}$. The vessel is assumed to be much larger than the manipulator, such that dynamic coupling is one-way.

The control objective, illustrated in Fig.~\ref{fig:problem_visualisation}, is to maintain a resilient communication channel with a shore-based station. This is done by tracking the target pose in orientation and vertical position by minimising the rotational error and compensating for vessel heave, while rejecting the perturbations $\tau_d$ in~\eqref{eq:disturbance_decomposition}. The rotational error is defined following~\citep{campa_2009_pose} as
\begin{equation}
    e_R(\tilde{R}) \triangleq
    \frac{1}{2}
    \begin{bmatrix}
        \tilde{r}_{32} - \tilde{r}_{23} \\
        \tilde{r}_{13} - \tilde{r}_{31} \\
        \tilde{r}_{21} - \tilde{r}_{12}
    \end{bmatrix}
    \in \mathbb{R}^3
    \label{eq:rotation_error_vector}
\end{equation}
where $\tilde{r}_{ij}$ is the $(i,j)$-th entry of the rotation error matrix
\begin{equation}
    \tilde{R} \triangleq R_\mathrm{tgt} R^\mathrm{T} \in \mathrm{SO}(3)
    \label{eq:rotation_error_matrix}
\end{equation}
with $R$ and $R_\mathrm{tgt}$ representing the current and desired antenna orientations, respectively.

The vertical position error is defined in the world frame as
\begin{equation}
    e_z \triangleq z^\mathrm{W} - z_\mathrm{tgt}^\mathrm{W} 
    \label{eq:vertical_position_error}
\end{equation}
where $z_\mathrm{tgt}^\mathrm{W}$ and $z^\mathrm{W}$ denote the desired and current antenna heights, respectively. In the simulation scenarios, a fixed world frame heave target is used.

%% file: controlDesign.tex

\begin{figure}[H]
    \centering
    \includegraphics[width=\linewidth]{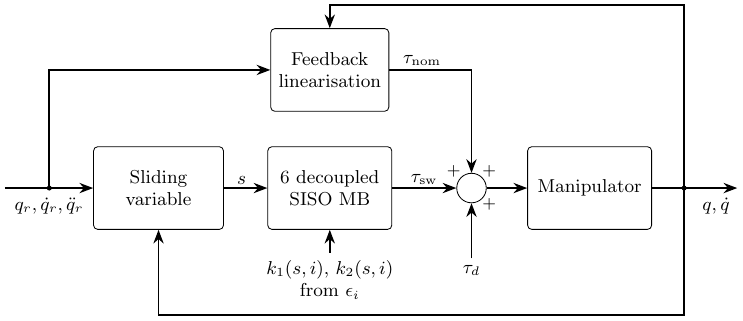}
    \caption{Proposed control architecture.}
    \label{fig:control_architecture}
\end{figure}

This section presents the proposed resilience scheme for ship-to-base communication, illustrated in Figure~\ref{fig:control_architecture}. Resilient antenna stabilisation is achieved using the multi-barrier super-twisting sliding mode controller (MB-STSMC), which is designed to maintain prescribed tracking performance despite environmental disturbances and model uncertainties.
\subsection{Resilience scheme}

Using nominal model parameters, feedback linearisation is employed to decouple the manipulator dynamics into six perturbed integrators. 
The resulting 
joint dynamics are controlled independently using the MB-STSMC, whose nested performance barriers provide resilience against perturbations that may violate inner tracking bounds while constraining the tracking error within the prescribed outermost bound, subject to the stated assumptions.

The reference trajectory $(q_r,\dot{q}_r,\ddot{q}_r)$ is generated by a standard inverse-kinematics null-space solution that guarantees $q_r \in C^2$. To decouple trajectory tracking from pose estimation, reference generation assumes knowledge of the world poses of both the vessel and the shore station.

\subsection{Application to the Manipulator} \label{sec:mani_application}
The manipulator dynamics in~\eqref{eq:mani_dynamics} reads:
\begin{equation}
    \ddot{q} = D^{-1}\left(\tau + \tau_d - C\,\dot{q} - G\right)
\end{equation}
The goal is to stabilise the error dynamics by ensuring each joint tracks the reference. The joint tracking error vector $e\in\mathbb{R}^6$ is therefore defined as:
\begin{equation}
    e \triangleq q - q_r
\end{equation}
where $q_r\in\mathbb{R}^6$ is the joint reference vector.

Defining the vector sliding variable $s\in\mathbb{R}^6$ as:
\begin{equation}
    s \triangleq \dot e + \Lambda e,
    \qquad
    \Lambda\triangleq\mathrm{diag}(\lambda_1,\ldots,\lambda_6),
    \quad \lambda_i>0
\end{equation}
results in the sliding variable dynamics:
\begin{equation}
    \dot{s}
    =
    D^{-1}\left(\tau + \tau_d - C\,\dot{q} - G\right)
    - \ddot{q}_r
    + \Lambda\left(\dot{q} - \dot{q}_r\right)
\end{equation}
The control input $\tau\in\mathbb{R}^6$ is then selected as a combination of feedback linearisation and a robust switching term:
\begin{equation}
    \begin{aligned}
        \tau &= \tau_{\mathrm{nom}} + \tau_{\mathrm{sw}} \\
          &= \underbrace{\hat{C}\,\dot{q} + \hat{G} + \hat{D}\left(\ddot{q}_r - \Lambda\left(\dot{q} - \dot{q}_r\right)\right)}_{\tau_{\mathrm{nom}}} + \underbrace{\hat{D}\,u_{\mathrm{sw}}}_{\tau_{\mathrm{sw}}}
    \end{aligned}
    \label{eq:3:control_law}
\end{equation}
Using this law, the robot joint dynamics are decoupled, with each joint being described by a scalar perturbed integrator:
\begin{equation}
    \dot s = u_{\mathrm{sw}} + d(t)
    \label{eq:robot_perturbed_integrator}
\end{equation}
where $d(t)$ lumps all remaining additive effects, including external perturbations, residual friction, model mismatch, and off-diagonal cross-coupling and most importantly, the effect of faults. 
Faults are restricted to matched additive torques with bounded magnitude and rate, so that $d$ satisfies:
\begin{assum}
    Each of the lumped disturbances $d(t)$ has a bounded time derivative, i.e. $\exists\,\Delta>0$ s.t. $\forall t\in\mathbb{R}^+$, $\|\dot{d}(t)\|\leq\Delta$, where each $\Delta$ is a priori unknown.
    \label{ass:tau_d}
\end{assum}

The following assumption accounts for residual multiplicative input-gain uncertainty, relative to~\eqref{eq:robot_perturbed_integrator}, arising from a potential mismatch between $\hat{D}$ and $D$.

\begin{assum}
    The inertia matrix $\hat{D}$ is identified sufficiently well such that the effective joint input gains remain positive and bounded, and the residual disturbances normalised by these gains satisfy Assumption \ref{ass:tau_d}.
    \label{ass:input_gain}
\end{assum}


\subsection{Super-twisting control of a perturbed integrator} 
Consider the perturbed integrator from \eqref{eq:robot_perturbed_integrator}.
Using the notation $\lfloor s \rceil^\alpha \triangleq |s|^\alpha\,\mathrm{sgn}(s)$, the STSMC control law is expressed as:
\begin{subequations}
    \label{eq:control_law_perturbed}
    \begin{align}
        u &= -k_1 \lfloor s \rceil^\frac{1}{2} + v \\
        \dot{v} &= -k_2 \lfloor s \rceil^0
    \end{align}
\end{subequations}
where $k_1, k_2\in \mathbb{R}^{*+}$. The closed-loop system, whose solutions are understood in the sense of Filippov, is then:
\begin{subequations}
    \label{eq:closedloop}
    \begin{align}
        \dot{s} &= -k_1 \lfloor s \rceil^\frac{1}{2} + \phi \\
        \dot{\phi} &= -k_2 \lfloor s \rceil^0 + \delta(t)
    \end{align}
\end{subequations}
with $\delta(t) \triangleq \dot{d}(t)$. It has been shown \citep{moreno_2012_lyapunov} that appropriate selection of the positive gains $k_1,k_2$ ensure finite-time stability of the origin of \eqref{eq:closedloop}.


\subsection{Multi-layered Sliding Mode Control}

\begin{figure}[t]
    \centering
    \includegraphics[width=\linewidth]{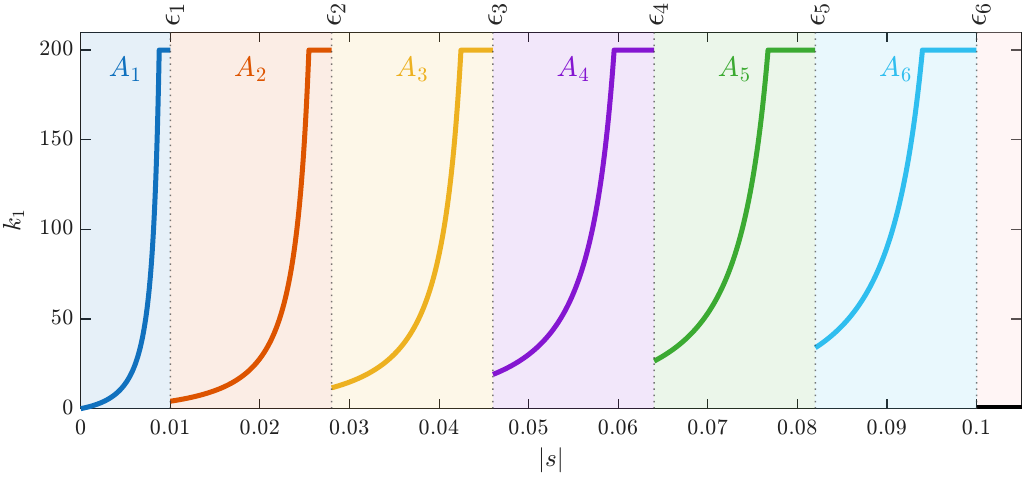}
    \caption{MB gain selection where $N=6$, $\epsilon_1=0.01$ and $\epsilon_N=0.1$. The gain is clamped at $k_1=200$ for visual clarity.}
    \label{fig:MB_detail}
\end{figure}

The multi-barrier (MB) formulation uses the control law in~\eqref{eq:control_law_perturbed}, but replaces the fixed gains with adaptive gains. To achieve this, it relies on a sequence of $N$ nested barriers:
\begin{equation}
    \epsilon_0\triangleq 0,
    \qquad
    0<\epsilon_1<\epsilon_2<\cdots<\epsilon_N
\end{equation}
where $\epsilon_N$ defines the maximum allowed deviation of the sliding variable, while the inner layers impose progressively strict performance bounds. A layer $i$ is active when:
\begin{equation}
    \epsilon_{i-1}\leq |s|<\epsilon_i
\end{equation}
Based on the position of the sliding variable in the active layer, the gains are computed as:
\begin{equation}
    k_1(s,i) = \frac{|s|}{(\epsilon_i-|s|)^{\frac{1}{2}+1}},
    \qquad
    k_2(s,i)=k_1(s,i)^2
    \label{eq:2:barrier_equation}
\end{equation}
Using this law, it is seen that as $|s|\rightarrow\,\epsilon_i^-$, the gain $k_1\rightarrow\infty$. In~\citep{vie_MB}, this is shown to result in uniform ultimate boundedness (UUB) of the sliding variable across the defined layers in continuous time. Moreover, the barrier-layer invariance property guarantees that each layer is forward invariant:
\begin{equation}
    |s(0)|<\epsilon_i
    \quad\Rightarrow\quad
    |s(t)|<\epsilon_i,\ \forall t\geq 0
    \label{eq:2:forward_invariance}.
\end{equation}
This invariance holds only in continuous time. For discrete-time implementation, inter-sample blindness may cause the sliding variable to cross barriers. The gain selection is visualised in Fig.~\ref{fig:MB_detail}.

\begin{rem}
    If the sliding variable exceeds the outermost barrier $\epsilon_N$, the gain is increased according to $k_{1,\mathrm{a}}=\exp(t_{\mathrm{a}})$, with $k_{2,\mathrm{a}}=k_{1,\mathrm{a}}^2$, where $t_{\mathrm{a}}$ denotes the time spent outside the admissible region. This mechanism ensures that, upon loss of confinement within the barrier structure, a sufficiently aggressive gain escalation is enforced to drive the system back toward the admissible region. Upon re-entry, the adaptive gain is reset.
\end{rem}

\begin{figure}[t]
    \centering
    \includegraphics[width=0.7\linewidth]{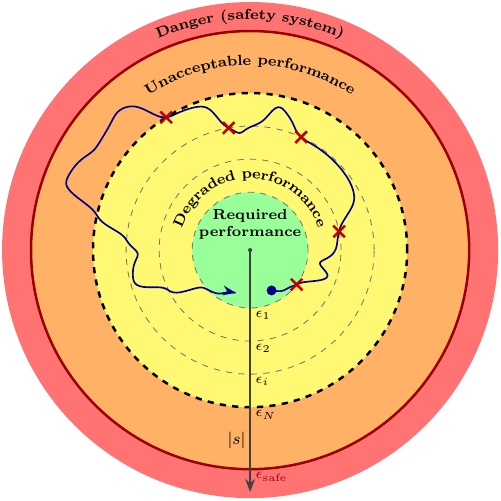}
    \caption{Conceptual schematic of the multi-barrier layers, with radial distance $|s|$, related to the performance regions of fault-tolerant control~\citep{blanke_2016_diagnosis}.}
    \label{fig:MB_regions}
\end{figure}

Through the choice of sliding manifold, e.g. $s=\dot e+\lambda e$ with $\lambda>0$, confinement of $s$ within layer $i$ implies the ultimate bound $|e|\leq\frac{\epsilon_i}{\lambda}$ on the tracking error. The barrier layers thus map directly onto performance requirements, as illustrated in Fig.~\ref{fig:MB_regions}, where red "$\times$" marks outward barrier crossings that occur between samples in discrete-time implementation. Condition $\vert s \vert \geq \epsilon_{\mathrm{safe}}>\epsilon_N$ denotes violation of safe operation, which is outside the scope of this work.

%% file: results.tex

\subsection{Simulation Environment}
The motion of the ship on which the manipulator is mounted is generated using the Marine Systems Simulator (MSS)~\citep{fossen_2004_mss}, where the RAO model of an $82.8\,\mathrm{m}$ supply vessel was considered. The marine craft modelling framework is described in~\citet{fossen_2011_handbook}. Sea-state conditions are defined according to the WMO international code for sea-state reporting~\citep{WMO2019}.

\subsection{Ship motion and environmental disturbances}

The vessel motion is accounted for using a simplified effective-gravity approach, assuming that base acceleration acts uniformly over the manipulator. The effective gravity is defined as
\begin{equation}
    g^\mathrm{B}_{\mathrm{eff}}
    \triangleq g^\mathrm{B}-a^\mathrm{B},
    \qquad
    g^\mathrm{B} \triangleq(R^\mathrm{W}_\mathrm{B})^\mathrm{T}g^\mathrm{W}
\end{equation}
where \(a^\mathrm{B}\) is the base acceleration derived from the vessel's translational and angular motion, obtained from the MSS, using standard rigid-body kinematics~\citep{fossen_2011_handbook}. The resulting wave-induced torque is
\begin{equation}
    \tau_{\mathrm{wave}}
    \triangleq G(q,g^\mathrm{B})-G(q,g^\mathrm{B}_{\mathrm{eff}})
\end{equation}
where $G(q,g)$ denotes the gravity torque vector evaluated at configuration $q$ for a given gravity vector $g$. The robot joint friction is included using a viscous-Coulomb model:
\begin{equation}
    \tau_{\mathrm{fric}}
    \triangleq -B_v\dot{q}-F_c\tanh(k\dot{q}),
\end{equation}
where $B_v$ and $F_c$ are diagonal friction coefficient matrices and $k>0$ regulates the steepness of the Coulomb friction model. Finally, wind is modelled as an external force acting on the end effector, mapped to joint torques through:
\begin{equation}
        \tau_{\mathrm{wind}} \triangleq J_p(q)^\mathrm{T}F^\mathrm{B}_{\mathrm{wind}},
        \quad
        F_0 = \frac{1}{2}\rho V^2 C_D A
\end{equation}
with $J_p(q)\in\mathbb{R}^{3\times6}$ the translational Jacobian, $F^\mathrm{B}_{\mathrm{wind}}\in\mathbb{R}^3$ the wind force expressed in the manipulator base frame, based on force magnitude $F_0$, $\rho$ the air density, $V$ the wind speed, $C_D$ the drag coefficient and $A$ the reference area. The parameters $C_D$ and $A$ are assumed constant and independent of configuration. The wind speed $V$ is selected using a sea-state-dependent mapping based on the Beaufort wind scale~\citep{beaufort_scale}. The wind force comprises a steady component and two sinusoidal gust components, with its direction specified in $\mathcal{F}_\mathrm{W}$ and transformed into $\mathcal{F}_\mathrm{B}$.

The controllers are implemented using the eigenvalue-matching discretisation scheme proposed by~\citep{koch_2019_discretetime} and further applied in~\citep{vie_2026_discrete} for all simulation results.

\subsection{Simulation scenarios and performance metrics} \label{sec:sim_scenarios}
The performance of the proposed scheme is evaluated in three configurations:
\begin{enumerate}
    \item Multi-barrier with $N=6$, $\epsilon_1=0.001$, $\epsilon_N=0.1$.
    \item Single-barrier (relaxed) with $\epsilon=\epsilon_N$.
    \item Single barrier (performance) with $\epsilon=\epsilon_1$.
\end{enumerate}
The simulation scenario is based on sea state 3, which is the highest sea state for which full heave compensation can be achieved without kinematic saturation. The system is first subjected to the environmental disturbances described by~\eqref{eq:disturbance_decomposition}, followed by a sudden perturbation introduced at $t=10\,\mathrm{s}$. This fault condition is introduced to evaluate the robustness of the controller against sudden, unexpected loads under discrete-time sampling. It represents a cable snag or tether entanglement, modelled as a sustained external force $F^\mathrm{B}_\mathrm{fault} = [100, -100, 50]^\mathrm{T}\,\mathrm{N}$ applied at the end effector. The force is introduced through a linear ramp with a rise time of $50\,\mathrm{ms}$, ensuring a bounded time derivative consistent with Assumption~\ref{ass:tau_d}.

This comparison evaluates whether the nested barriers provide an advantage over either fixed performance bound under the same disturbance conditions.
The results are quantified according to:
\begin{itemize}
    \item $t_{\mathrm{escape}}$: Time spent outside barrier regime.
    \item $e_{R}$: Maximum and average pointing error.
    \item $e_{z}$: Maximum and average heave compensation error.
\end{itemize}

\subsection{Results}
\begin{figure}[t]
    \centering
    \includegraphics[width=1\linewidth]{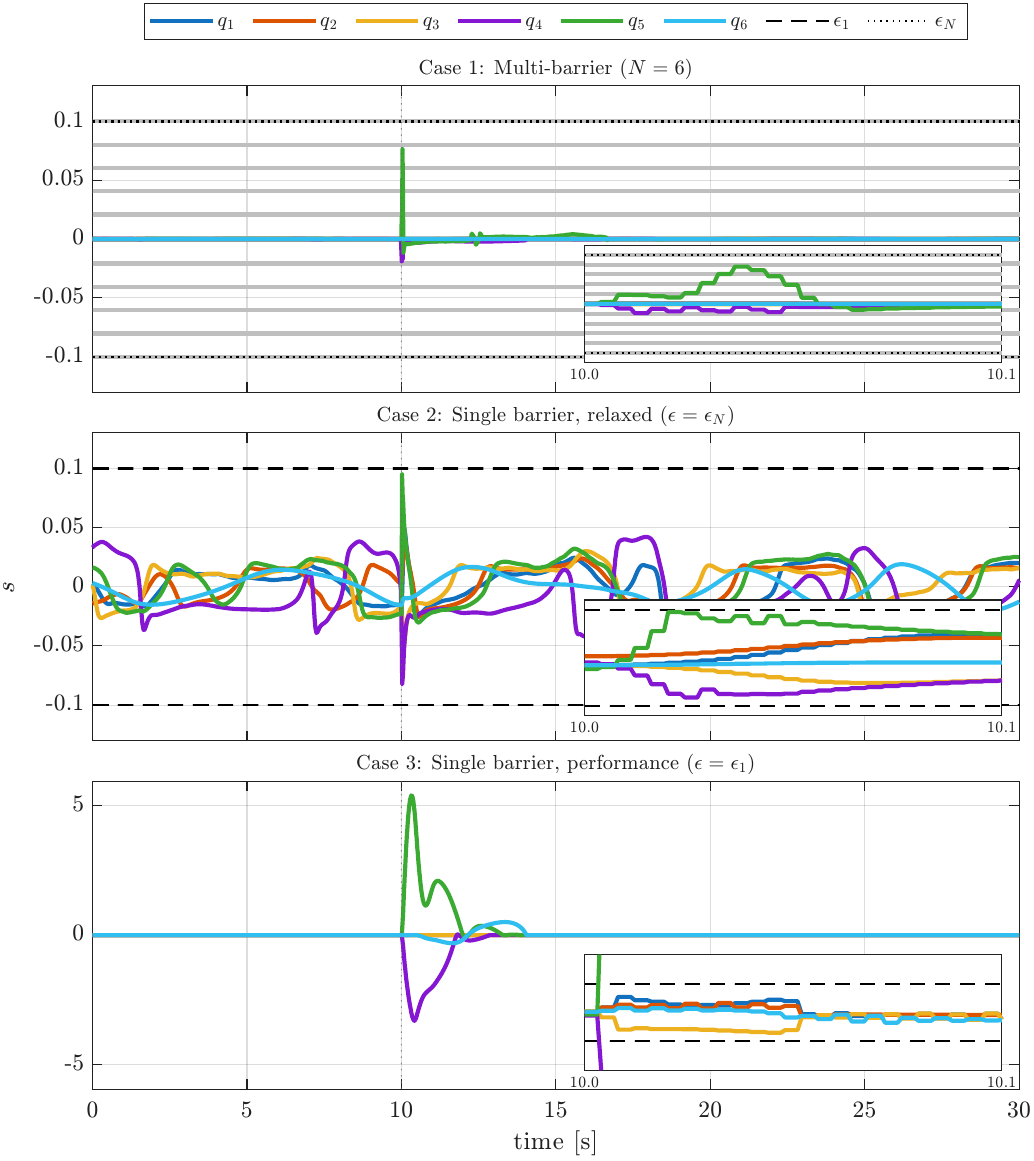}
    \caption{Sliding variables $s_1$-$s_6$ for the multi-layer (top), relaxed single layer (middle) and performance single layer (bottom) variants of the proposed scheme. The barriers are shown as dashed ($\epsilon_1$) and dotted ($\epsilon_N$) lines.}
    \label{fig:scenarios_sliding}
\end{figure}

\begin{figure}[t]
    \centering
    \includegraphics[width=1\linewidth]{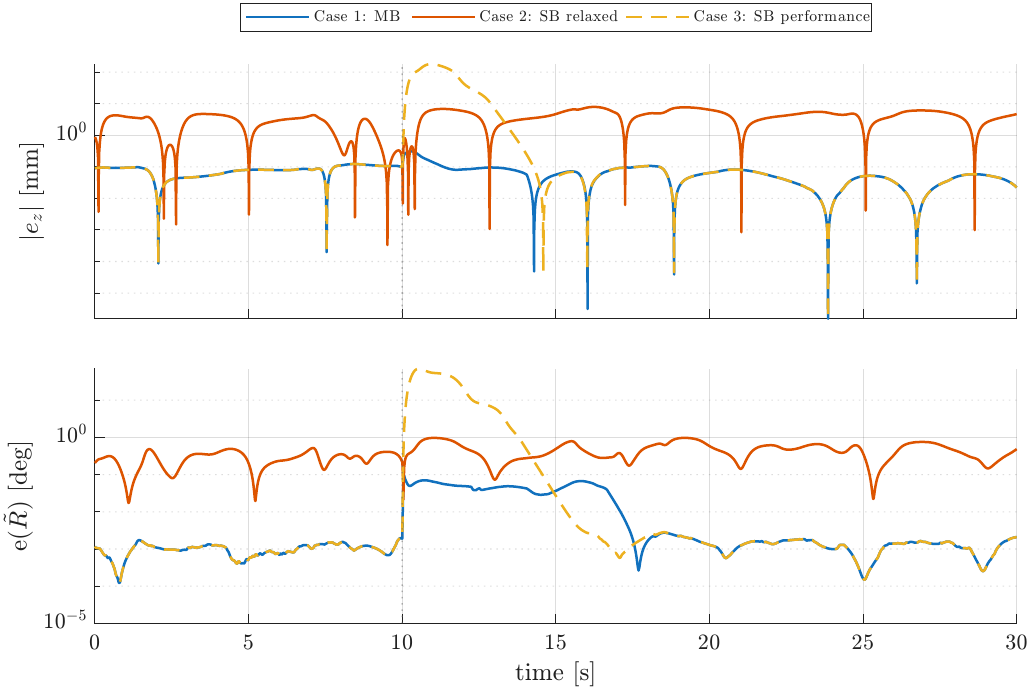}
    \caption{Heave error (absolute value) and pointing error for the three configurations on a logarithmic scale.}
    \label{fig:scenarios_error}
\end{figure}

Fig. \ref{fig:scenarios_sliding} shows the time evolution of the sliding variables for each joint for all scenarios. The zoomed plots show the first $0.1\,\mathrm{s}$ after the fault. The multi-barrier controller quickly returns to the inner barrier, the relaxed single barrier remains bounded by $\epsilon_N$ but never settles tightly, and the performance single barrier variant leaves its barrier and takes several seconds to recover. A similar conclusion can be drawn from Fig. \ref{fig:scenarios_error}, where the heave and pointing error are illustrated. Specifically, it can be seen that before the fault, the multi-barrier and performance single-barrier controllers have similar performance. After the fault at $t=10\,\mathrm{s}$, the performance single barrier deviates by several orders of magnitude, while the multi-barrier controller remains close to its pre-fault level. The relaxed single barrier has a higher error throughout. The results are also confirmed in terms of the selected metrics reported in Table \ref{tab:scenario_metrics}.

\begin{table}[t]
    \centering
    \caption{Performance after the fault ($t \geq 10\,\mathrm{s}$).}
    \label{tab:scenario_metrics}
    \resizebox{\linewidth}{!}{%
    \begin{tabular}{lccccc}
        \toprule
         & $t_{\mathrm{escape}}$ & \multicolumn{2}{c}{$e_R$ [deg]} & \multicolumn{2}{c}{$e_z$ [mm]} \\
        \cmidrule(lr){3-4} \cmidrule(lr){5-6}
        Controller & [s] & max & mean & max & mean \\
        \midrule
        Multi-barrier        & 0.00 & 0.098 & 0.018 & 0.31 & 0.06 \\
        Single (relaxed)     & 0.00 & 0.970 & 0.477 & 7.92 & 4.44 \\
        Single (performance) & 4.10 & 69.80 & 5.159 & 180.57 & 12.64 \\
        \bottomrule
    \end{tabular}%
    }
\end{table}


%% file: conclusions.tex
A multi-layered sliding mode passive fault-tolerant control was presented in this work for addressing the problem of ship-to-base resilient communication via antenna robust stabilisation and alignment. The antenna was mounted on a 6 DOF robotic manipulator on board the vessel, whereas faults and environmental disturbances as well as parametric model uncertainty were described as rate-bounded perturbations. The proposed scheme was shown to preserve the end-effector pose error within the prescribed performance bounds demonstrating superior rejection of fault-induced perturbations compared to single-layered adaptive sliding mode control variants. Future work will pursue implicit discretisation of the presented controller and experimental verification of its fault-tolerant capabilities.

%% file: resilientCommSafeProc2026.bib
@book{fossen_2011_handbook,
  author    = {Fossen, Thor I.},
  title     = {Handbook of Marine Craft Hydrodynamics and Motion Control},
  publisher = {Wiley},
  address   = {Chichester, West Sussex},
  year      = {2011}
}

@misc{fossen_2004_mss,
  author = {Fossen, Thor I. and Perez, Tristan},
  title  = {Marine Systems Simulator ({MSS})},
  url    = {https://github.com/cybergalactic/MSS},
  year   = {2004}
}

@misc{beaufort_scale,
  author = {{RMetS}},
  title  = {The {B}eaufort wind scale},
  url    = {https://www.rmets.org/metmatters/beaufort-wind-scale},
  year   = {2026}
}

@article{wang_2021_attitude,
  author  = {Wang, Yunlong and Soltani, Mohsen and Hussain, Dil Muhammad Akbar},
  title   = {Attitude stabilization of marine satellite tracking antenna using model predictive control},
  journal = {IFAC Journal of Systems and Control},
  volume  = {17},
  pages   = {100173},
  year    = {2021},
  doi     = {10.1016/j.ifacsc.2021.100173}
}

@article{wang_2018_wireless,
  author  = {Wang, Jue and Zhou, Haifeng and Li, Ye and Sun, Qiang and Wu, Yongpeng and Jin, Shi and Quek, Tony Q. S. and Xu, Chen},
  title   = {Wireless channel models for maritime communications},
  journal = {IEEE Access},
  volume  = {6},
  pages   = {68070--68088},
  year    = {2018},
  doi     = {10.1109/ACCESS.2018.2879902}
}

@article{vie2025extended,
  title={Extended Time Varying Multi-Cluster Fluctuating Two-Ray Fading Model for Maritime Environment},
  author={Vi{\'e}, Antoine Thibault and Galeazzi, Roberto and Papagergiou, Dimitrios},
  journal={IFAC-PapersOnLine},
  volume={59},
  number={22},
  pages={734--739},
  year={2025},
  publisher={Elsevier}
}

@article{utkin_1977_variable,
  author  = {Utkin, Vadim},
  title   = {Variable structure systems with sliding modes},
  journal = {IEEE Transactions on Automatic Control},
  volume  = {22},
  number  = {2},
  pages   = {212--222},
  year    = {1977}
}

@article{obeid_2018_barrier,
  author  = {Obeid, Hussein and Fridman, Leonid M. and Laghrouche, Salah and Harmouche, Mohamed},
  title   = {Barrier function-based adaptive sliding mode control},
  journal = {Automatica},
  volume  = {93},
  pages   = {540--544},
  year    = {2018},
  doi     = {10.1016/j.automatica.2018.03.078}
}

@article{obeid_2020_barrier,
  author  = {Obeid, Hussein and Laghrouche, Salah and Fridman, Leonid and Chitour, Yacine and Harmouche, Mohamed},
  title   = {Barrier function-based adaptive super-twisting controller},
  journal = {IEEE Transactions on Automatic Control},
  volume  = {65},
  number  = {11},
  pages   = {4928--4933},
  year    = {2020},
  doi     = {10.1109/TAC.2020.2974390}
}

@article{moreno_2014_strict,
  author  = {Moreno, Jaime A.},
  title   = {On strict {L}yapunov functions for some non-homogeneous super-twisting algorithms},
  journal = {Journal of the Franklin Institute},
  volume  = {351},
  number  = {4},
  pages   = {1902--1919},
  year    = {2014},
  doi     = {10.1016/j.jfranklin.2013.09.019}
}

@inproceedings{vie_MB,
  author    = {Vi{\'e}, Antoine Thibault and Fridman, Leonid and Galeazzi, Roberto and Papageorgiou, Dimitrios},
  title     = {Multi-layer barrier function-based adaptive super-twisting controller},
  booktitle = {2025 IEEE 64th Conference on Decision and Control (CDC)},
  pages     = {7932--7938},
  publisher = {IEEE},
  year      = {2025},
  doi       = {10.1109/CDC57313.2025.11312295}
}

@article{koch_2019_discretetime,
  author  = {Koch, Stefan and Reichhartinger, Markus},
  title   = {Discrete-time equivalents of the super-twisting algorithm},
  journal = {Automatica},
  volume  = {107},
  pages   = {190--199},
  year    = {2019},
  doi     = {10.1016/j.automatica.2019.05.040},
  url     = {https://www.sciencedirect.com/science/article/pii/S0005109819302596}
}

@inproceedings{eisenzopf2021adaptive,
  author    = {Eisenzopf, L. and Watermann, L. and Koch, S. and Reichhartinger, M. and Reger, J. and Horn, M.},
  title     = {Adaptive Gain Super-Twisting-Algorithm: Design and Discretization},
  booktitle = {2021 60th IEEE Conference on Decision and Control (CDC)},
  pages     = {6415--6420},
  year      = {2021},
  doi       = {10.1109/CDC45484.2021.9683304},
  url       = {https://ieeexplore.ieee.org/document/9683304},
  issn      = {2576-2370}
}

@article{papageorgiou2022behaviour,
  author  = {Papageorgiou, D. and Edwards, C.},
  title   = {On the behaviour of under-tuned super-twisting sliding mode control loops},
  journal = {Automatica},
  volume  = {135},
  pages   = {109983},
  year    = {2022},
  doi     = {10.1016/j.automatica.2021.109983}
}

@article{gonzalez2024final,
  author  = {Gonz{\'a}lez, A. and Ovalle, L. and Fridman, L.},
  title   = {Final Set Adjustment in Barrier Function Adaptation Exploiting Properties of Signed Power-Based Controllers},
  journal = {IEEE Transactions on Automatic Control},
  volume  = {PP},
  pages   = {1--6},
  year    = {2024},
  doi     = {10.1109/TAC.2024.3367285}
}

@article{cruz_ancona_2021_barrier,
  author  = {Cruz-Ancona, C. D. and Estrada, M. A. and Fridman, L.},
  title   = {Barrier Function-Based Adaptive Lyapunov Redesign for Systems Without A Priori Bounded Perturbations},
  journal = {IEEE Transactions on Automatic Control},
  volume  = {67},
  number  = {8},
  pages   = {3851--3862},
  year    = {2022}
}

@article{moreno_2015_adaptive,
  author  = {Moreno, J. and Negrete, D. and Torres-Gonz{\'a}lez, V. and Fridman, L.},
  title   = {Adaptive Continuous Twisting Algorithm},
  journal = {International Journal of Control},
  pages   = {1--15},
  year    = {2015},
  doi     = {10.1080/00207179.2015.1116713}
}

@incollection{moreno_2012_lyapunov,
  author    = {Moreno, J. A.},
  title     = {Lyapunov Approach for Analysis and Design of Second Order Sliding Mode Algorithms},
  booktitle = {Sliding Modes after the First Decade of the 21st Century: State of the Art},
  editor    = {Fridman, L. and Moreno, J. and Iriarte, R.},
  series    = {Lecture Notes in Control and Information Sciences},
  pages     = {113--149},
  publisher = {Springer},
  address   = {Berlin, Heidelberg},
  year      = {2012},
  doi       = {10.1007/978-3-642-22164-4}
}

@article{ovalle_2025_noise,
  author  = {Ovalle, L. and Gonzalez, A. and Fridman, L. and Laghrouche, S. and Obeid, H.},
  title   = {Analysis of Barrier Function Based Adaptive Sliding Mode Control in the Presence of Deterministic Noise},
  journal = {Automatica},
  volume  = {171},
  pages   = {111946},
  year    = {2025}
}

@inproceedings{shtessel2010super,
  author    = {Shtessel, Y. and Moreno, J. and Plestan, F. and Fridman, L. and Poznyak, A.},
  title     = {Super-Twisting Adaptive Sliding Mode Control: A Lyapunov Design},
  booktitle = {Proceedings of the 49th IEEE Conference on Decision and Control},
  pages     = {5109--5113},
  year      = {2010},
  doi       = {10.1109/CDC.2010.5717908}
}

@article{utkin_2013_adaptive,
  author  = {Utkin, V. I. and Poznyak, A. S.},
  title   = {Adaptive Sliding Mode Control with Application to Super-Twist Algorithm: Equivalent Control Method},
  journal = {Automatica},
  volume  = {49},
  number  = {1},
  pages   = {39--47},
  year    = {2013},
  doi     = {10.1016/j.automatica.2012.09.008}
}

@inproceedings{vie_2026_discrete,
  author = {Vi{\'e}, A. T. and Fridman, L. and Galeazzi, R. and Papageorgiou, D.},
  title  = {Multi-layer barrier adaptation of the discrete-time super-twisting algorithm},
  year   = {2026}
}

@article{laghrouche_2021_barrier,
  author  = {Laghrouche, S. and Harmouche, M. and Chitour, Y. and Obeid, H. and Fridman, L. M.},
  title   = {Barrier Function-Based Adaptive Higher Order Sliding Mode Controllers},
  journal = {Automatica},
  volume  = {123},
  pages   = {109355},
  year    = {2021},
  doi     = {10.1016/j.automatica.2020.109355}
}

@book{blanke_2016_diagnosis,
  author    = {Blanke, Mogens and Kinnaert, Michel and Lunze, Jan and Staroswiecki, Marcel},
  title     = {Diagnosis and Fault-Tolerant Control},
  edition   = {3},
  publisher = {Springer},
  address   = {Berlin, Heidelberg},
  year      = {2016},
  doi       = {10.1007/978-3-662-47943-8}
}

@inproceedings{campa_2009_pose,
    author    = {Campa, Ricardo and de la Torre, Hussein},
    title     = {Pose Control of Robot Manipulators Using Different Orientation Representations: A Comparative Review},
    booktitle = {Proceedings of the 2009 American Control Conference},
    pages     = {2855--2860},
    year      = {2009},
    doi       = {10.1109/ACC.2009.5160254}
}

@book{WMO2019,
  author    = {{World Meteorological Organization}},
  title     = {Manual on Codes: International Codes, Volume I.1},
  year      = {2019},
  publisher = {World Meteorological Organization},
  address   = {Geneva},
  number    = {306}
}

@article{estrada2024passive,
  title={Passive fault-tolerant control via sliding-mode-based Lyapunov redesign},
  author={Estrada, Manuel A and Fridman, Leonid and Moreno, Jaime A},
  journal={IEEE Transactions on Automatic Control},
  volume={69},
  number={10},
  pages={6777--6788},
  year={2024},
  publisher={IEEE}
}
